\documentclass[conference]{IEEEtran}[conference]

\newcommand{\pdftitle}{Towards a Sustainable Age of Information Metric: Carbon Footprint of Real-Time Status Updates}

\usepackage[nolist,nohyperlinks]{acronym}
\usepackage{amsmath}
\usepackage{breqn}
\usepackage{balance}
\usepackage{cite}
\usepackage{framed}
\usepackage{siunitx}
\usepackage{soul}
\usepackage{multirow}
\usepackage{etoolbox}

\usepackage{blindtext}

\usepackage{standalone}

\usepackage{tikz}
\usepackage{pgfplots}
\usetikzlibrary{spy}
\usepgfplotslibrary{groupplots}
\usetikzlibrary{backgrounds}

\usepackage{graphicx}
\usepackage{subcaption}
\usepackage{color}
\usepackage{booktabs}
\usepackage{float}
\usepackage[table,xcdraw]{xcolor}

\graphicspath{{./images/}}

\usepackage{url}
\usepackage[bookmarks=false]{hyperref}
\hypersetup{
  backref,
  hidelinks,
  colorlinks=false,
  breaklinks=true,
  bookmarksopen=false,
  pdftitle=\pdftitle,
  pdfauthor={Author}
}

\newtoggle{authorcomment}
\toggletrue{authorcomment}

\newcommand\copyrighttext{%
  \footnotesize \textcopyright 2026 IEEE. Personal use of this material is permitted.
  Permission from IEEE must be obtained for all other uses, in any current or future
  media, including reprinting/republishing this material for advertising or promotional
  purposes, creating new collective works, for resale or redistribution to servers or
  lists, or reuse of any copyrighted component of this work in other works. This paper was accepted for presentation at ICC 2026.
  }
\newcommand\copyrightnotice{%
\begin{tikzpicture}[remember picture,overlay]
\node[anchor=north,yshift=-10pt] at (current page.north) {\fbox{\parbox{\dimexpr\textwidth-\fboxsep-\fboxrule\relax}{\copyrighttext}}};
\end{tikzpicture}%
}

\begin{document}
\bstctlcite{IEEEexample:BSTcontrol}

\title{\pdftitle}

\author{
    \IEEEauthorblockN{Shih-Kai Chou\IEEEauthorrefmark{1},
                      Maice Costa\IEEEauthorrefmark{2},
                      Mihael Mohor\v{c}i\v{c}\IEEEauthorrefmark{1}, and
                      Jernej Hribar\IEEEauthorrefmark{1}}
    \IEEEauthorblockA{\IEEEauthorrefmark{1}Jožef Stefan Institute, Ljubljana, Slovenia,\\
                      \IEEEauthorrefmark{2}Nexcepta Inc., Gaithersburg MD 20878, USA,\\
                      Email: \{shih-kai.chou, mihael.mohorcic, jernej.hribar\}@ijs.si, mcosta@nexcepta.com }
}

\maketitle

\begin{acronym}[MACHU]
  \acro{iot}[IoT]{Internet of Things}
  \acro{iiot}[IIoT]{Industrial Internet of Things}
  \acro{cr}[CR]{Cognitive Radio}
  \acro{ofdm}[OFDM]{orthogonal frequency-division multiplexing}
  \acro{ofdma}[OFDMA]{orthogonal frequency-division multiple access}
  \acro{scfdma}[SC-FDMA]{single carrier frequency division multiple access}
  \acro{rbi}[RBI]{ Research Brazil Ireland}
  \acro{rfic}[RFIC]{radio frequency integrated circuit}
  \acro{sdr}[SDR]{Software Defined Radio}
  \acro{sdn}[SDN]{Software Defined Networking}
  \acro{su}[SU]{Secondary User}
  \acro{ra}[RA]{Resource Allocation}
  \acro{qos}[QoS]{quality of service}
  \acro{usrp}[USRP]{Universal Software Radio Peripheral}
  \acro{mno}[MNO]{Mobile Network Operator}
  \acro{mnos}[MNOs]{Mobile Network Operators}
  \acro{gsm}[GSM]{Global System for Mobile communications}
  \acro{tdma}[TDMA]{Time-Division Multiple Access}
  \acro{fdma}[FDMA]{Frequency-Division Multiple Access}
  \acro{gprs}[GPRS]{General Packet Radio Service}
  \acro{msc}[MSC]{Mobile Switching Centre}
  \acro{bsc}[BSC]{Base Station Controller}
  \acro{umts}[UMTS]{universal mobile telecommunications system}
  \acro{Wcdma}[WCDMA]{Wide-band code division multiple access}
  \acro{wcdma}[WCDMA]{wide-band code division multiple access}
  \acro{cdma}[CDMA]{code division multiple access}
  \acro{lte}[LTE]{Long Term Evolution}
  \acro{papr}[PAPR]{peak-to-average power rating}
  \acro{hn}[HetNet]{heterogeneous networks}
  \acro{phy}[PHY]{physical layer}
  \acro{mac}[MAC]{medium access control}
  \acro{amc}[AMC]{adaptive modulation and coding}
  \acro{mimo}[MIMO]{multiple input multiple output}
  \acro{rats}[RATs]{radio access technologies}
  \acro{vni}[VNI]{visual networking index}
  \acro{rbs}[RB]{resource blocks}
  \acro{rb}[RB]{resource block}
  \acro{ue}[UE]{user equipment}
  \acro{cqi}[CQI]{Channel Quality Indicator}
  \acro{hd}[HD]{half-duplex}
  \acro{fd}[FD]{full-duplex}
  \acro{sic}[SIC]{self-interference cancellation}
  \acro{si}[SI]{self-interference}
  \acro{bs}[BS]{base station}
  \acro{fbmc}[FBMC]{Filter Bank Multi-Carrier}
  \acro{ufmc}[UFMC]{Universal Filtered Multi-Carrier}
  \acro{scm}[SCM]{Single Carrier Modulation}
  \acro{isi}[ISI]{inter-symbol interference}
  \acro{ftn}[FTN]{Faster-Than-Nyquist}
  \acro{m2m}[M2M]{machine-to-machine}
  \acro{mtc}[MTC]{machine type communication}
  \acro{mmw}[mmWave]{millimeter wave}
  \acro{bf}[BF]{beamforming}
  \acro{los}[LOS]{line-of-sight}
  \acro{nlos}[NLOS]{non line-of-sight}
  \acro{capex}[CAPEX]{capital expenditure}
  \acro{opex}[OPEX]{operational expenditure}
  \acro{ict}[ICT]{Information and Communications Technology}
  \acro{sp}[SP]{service providers}
  \acro{inp}[InP]{infrastructure providers}
  \acro{mvnp}[MVNP]{mobile virtual network provider}
  \acro{mvno}[MVNO]{mobile virtual network operator}
  \acro{nfv}[NFV]{network function virtualization}
  \acro{vnfs}[VNF]{virtual network functions}
  \acro{cran}[C-RAN]{Cloud Radio Access Network}
  \acro{bbu}[BBU]{baseband unit}
  \acro{bbus}[BBU]{baseband units}
  \acro{rrh}[RRH]{remote radio head}
  \acro{rrhs}[RRH]{Remote radio heads} 
  \acro{sfv}[SFV]{sensor function virtualization}
  \acro{wsn}[WSN]{wireless sensor networks} 
  \acro{bio}[BIO]{Bristol is open}
  \acro{vitro}[VITRO]{Virtualized dIstributed plaTfoRms of smart Objects}
  \acro{os}[OS]{operating system}
  \acro{www}[WWW]{world wide web}
  \acro{iotvn}[IoT-VN]{IoT virtual network}
  \acro{mems}[MEMS]{micro electro mechanical system}
  \acro{mec}[MEC]{Mobile edge computing}
  \acro{coap}[CoAP]{Constrained Application Protocol}
  \acro{vsn}[VSN]{Virtual sensor network}
  \acro{rest}[REST]{REpresentational State Transfer}
  \acro{aoi}[AoI]{Age of Information}
  \acro{lora}[LoRa\texttrademark]{Long Range}
  \acro{iot}[IoT]{Internet of Things}
  \acro{snr}[SNR]{Signal-to-Noise Ratio}
  \acro{cps}[CPS]{Cyber-Physical System}
  \acro{uav}[UAV]{Unmanned Aerial Vehicle}
  \acro{rfid}[RFID]{Radio-frequency identification}
  \acro{lpwan}[LPWAN]{Low-Power Wide-Area Network}
  \acro{lgfs}[LGFS]{Last Generated First Served}
  \acro{wsn}[WSN]{wireless sensor network} 
  \acro{lmmse}[LMMSE]{Linear Minimum Mean Square Error}
  \acro{rl}[RL]{Reinforcement Learning}
  \acro{nb-iot}[NB-IoT]{Narrowband IoT}
  \acro{lorawan}[LoRaWAN]{Long Range Wide Area Network}
  \acro{mdp}[MDP]{Markov Decision Process}
  \acro{ann}[ANN]{Artificial Neural Network}
  \acro{dqn}[DQN]{Deep Q-Network}
  \acro{mse}[MSE]{Mean Square Error}
  \acro{ml}[ML]{Machine Learning}
  \acro{cpu}[CPU]{Central Processing Unit}
  \acro{ddpg}[DDPG]{Deep Deterministic Policy Gradient}
  \acro{ai}[AI]{Artificial Intelligence}
  \acro{gp}[GP]{Gaussian Processes}
  \acro{drl}[DRL]{Deep Reinforcement Learning}
  \acro{mmse}[MMSE]{Minimum Mean Square Error}
  \acro{fnn}[FNN]{Feedforward Neural Network}
  \acro{eh}[EH]{Energy Harvesting}
  \acro{wpt}[WPT]{Wireless Power Transfer}
  \acro{dl}[DL]{Deep Learning}
  \acro{yolo}[YOLO]{You Only Look Once}
  \acro{mec}[MEC]{Mobile Edge Computing}
  \acro{marl}[MARL]{Multi-Agent Reinforcement Learning}
  \acro{aoi}[AoI]{Age of Information}
  \acro{cf}[CF]{Carbon Footprint}
  \acro{ci}[CI]{Carbon Intensity}
  \acro{fcfs}[FCFS]{First Come First Served}
  \acro{lcfs}[LCFS]{Last Come First Served}
  \acro{qos}[QoS]{Quality of Service}
  \acro{snr}[SNR]{Signal-to-noise Ratio}
  \acro{saoi}[SAoI]{Sustainable Age of Information}
  \acro{aqi}[AQI]{age and quality of Information}
  \acro{voi}[VoI]{Value of Information}
  \acro{qaoi}[QAoI]{Query Age of Information}
  \acro{aoii}[AoII]{Age of Incorrect Information}
  \acro{wsn}[WSN]{Wireless Sensor Network}
\end{acronym}

\begin{abstract}
The timeliness of collected information is essential for monitoring and control in data-driven intelligent infrastructures. It is typically quantified using the Age of Information (AoI) metric, which has been widely adopted to capture the freshness of information received in the form of status updates. While AoI-based metrics quantify how timely the collected information is, they largely overlook the environmental impact associated with frequent transmissions, specifically, the resulting Carbon Footprint (CF). 
To address this gap, we introduce a carbon-aware AoI framework. We first derive closed-form expressions for the average AoI under constrained CF budgets for the baseline $M/M/1$ and $M/M/1^*$ queuing models, assuming fixed Carbon Intensity (CI). We then extend the analysis by treating CI as a dynamic, time-varying parameter and solve the AoI minimization problem. 
Our results show that minimizing AoI does not inherently minimize CF, highlighting a clear trade-off between information freshness and environmental impact. CI variability further affects achievable AoI, indicating that sustainable operation requires joint optimization of CF budgets, Signal-to-noise Ratio (SNR), and transmission scheduling. This work lays the foundation for carbon-aware information freshness optimization in next-generation networks.

\end{abstract}

\acresetall

\begin{IEEEkeywords}
Age of Information, Carbon Footprint
\end{IEEEkeywords}

\copyrightnotice
\section{Introduction}
\label{sec:intro}

In the past decade, the \ac{aoi} metric~\cite{kaul2012real} has been introduced and gradually adopted to characterize information timeliness, offering valuable insights into the performance of real-time systems~\cite{yates2021age}. In the same period, we have witnessed the advent of increasingly data-driven and latency-sensitive intelligent infrastructures and applications, including autonomous systems, \ac{iiot}, remote healthcare, etc., for which \ac{aoi} provides a robust framework for optimizing information freshness. To address the needs of these emerging use cases, the original concept of \ac{aoi} has been extended to incorporate additional dimensions, aiming to characterize the diverse effects and relevance of timely information on overall system performance. For example, the Age of Incorrect Information (AoII)~\cite{maatouk2020age} combines \ac{aoi} with error penalty functions to assess the timeliness of status updates that contain new and correct information. Similarly, Effective AoI~\cite{yin2019only} characterizes \ac{aoi} from the user’s perspective by measuring only the freshness of information that is explicitly requested. Despite these advancements, none of the proposed metrics accounts for the environmental impact associated with transmitting fresh updates. To that end, in this paper, we investigate the connection between the concept of \ac{aoi} and sustainability, measured through the \ac{cf} produced by the system.





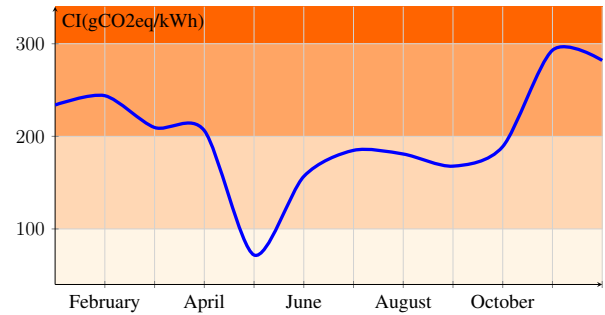
\begin{figure}
	\centering
	\large \begin{tikzpicture}
\definecolor{orange1}{RGB}{255, 245, 230} 
\definecolor{orange2}{RGB}{255, 215, 180} 
\definecolor{orange3}{RGB}{255, 165, 100} 
\definecolor{orange4}{RGB}{255, 100, 0}   

\large
\begin{axis}[
	axis lines=middle,
        ylabel=CI(gCO2eq/kWh),
	grid=both,
	major grid style={gray!35},
        ytick ={100, 200, 300},
        xtick ={0,1,2,3,4,5,6,7,8,9,10,11,12},
        xticklabels={, February, , April, , June, , August, , October, , },
        scale ticks above exponent={2},
        ymin=40,ymax=340,
	width=1.5\linewidth,
	height=0.85\linewidth,
	legend style={at={(0.35,0.95)}},
]

\begin{scope}[on background layer]
  \fill[orange1,opacity=10] ({rel axis cs:0,0}) rectangle ({rel axis cs:1.0,0.2});
    \fill[orange2,opacity=10] ({rel axis cs:0,.2}) rectangle ({rel axis cs:1.0,0.535});
     \fill[orange3,opacity=10] ({rel axis cs:0,.535}) rectangle ({rel axis cs:1.0,0.87});
     \fill[orange4,opacity=10] ({rel axis cs:0,.87}) rectangle ({rel axis cs:1.0,1.0});
\end{scope}
 
\addplot[line width=2pt, color=blue,  smooth, mark options={solid}]table[x=index,y=intensity,col sep=comma]{csv_data/intensity.csv};




\end{axis}

\end{tikzpicture}
         \caption{Annual Carbon Intensity (CI) in Slovenia (2024).}
	\label{fig:intro}
    \vspace{-15pt}
\end{figure}

Typically, the \ac{cf} of an \ac{ict} system is estimated by associating the energy consumed for data transmission with the corresponding \ac{ci} per unit of energy~\cite{CFreport}. 
The \ac{ci} is defined as the number of grams of CO$_2$-equivalent emissions produced per kilowatt-hour of energy consumed over a given time period, and is typically expressed in units of $\frac{gCO_{2}eq}{kWh}$. As such, \ac{ci} varies across time and geographic location. It depends on the mix of energy sources used in the local grid and is usually measured at the national, regional, or city level. As discussed in~\cite{trihinas2022towards}, energy sources such as biomass, coal, gas, and oil contribute to higher \ac{ci} values, whereas renewable (e.g., wind, solar, hydro, geothermal) and nuclear sources result in significantly lower emissions. To illustrate this, Fig.~\ref{fig:intro} shows the variation in \ac{ci} observed in Slovenia in 2024. Monthly differences are significant, with November emissions yielding up to three times the \ac{cf} of May. Such temporal variability indicates that assessing environmental impact solely by energy consumption may be inadequate.

The relationship between minimizing \ac{aoi} and improving energy efficiency, e.g., reducing energy consumption and average \ac{aoi},
has been explored in numerous studies, e.g.,~\cite{HUANG202329,xu2020info, zhang2023TWC,Hatami2021ToC, zhang2024TOC}. However, while minimizing energy consumption is a necessary step toward reducing the \ac{cf} of a system, it is not sufficient on its own. To truly minimize environmental impact, it is also necessary to consider the temporal and geographical context of energy usage, as the associated \ac{ci} can vary significantly over time and across regions. Hence, minimizing energy consumption does not necessarily minimize \ac{cf}, posing a key challenge for the design of sustainable communication systems.


To gain initial insights into newly identified problem, we investigate a system in which a single source transmits status updates to a monitor. The objective is to maintain information freshness while explicitly accounting for the \ac{cf} associated with transmissions. In this context, the key contributions of this paper are as follows:

\begin{itemize}
    \item We establish a link between information freshness and environmental impact by incorporating \ac{cf} into the evaluation of status update timeliness and show that minimizing \ac{aoi} does not necessarily yield minimal \ac{cf}. 

    \item We derive closed-form expressions for the average \ac{aoi} under \ac{cf} constraints for the baseline $M/M/1$ and $M/M/1^{*}$ models, showing that carbon constraints lower optimal system utilization and reveal a clear trade-off between timeliness and \ac{cf}.

    \item We model \ac{ci} as a time-varying parameter and solve the \ac{aoi} minimization problem under two edge cases. With both power and \ac{cf} constraints, the optimized \ac{aoi} in the $M/M/1$ model is about $1.45\times$ higher than in $M/M/1^{*}$ under low \ac{ci}. Under a minimum \ac{snr} constraint for \ac{qos}, the $M/M/1$ model shows nearly $2\times$ and $3\times$ higher \ac{aoi} at high and low \ac{ci}, respectively.

\end{itemize}

In short, our findings indicate that jointly considering \ac{cf} and \ac{aoi} of the system enables a more nuanced understanding of the trade-off between timeliness of collected information and sustainability of the system.
\section{Related work}
\label{sec:related}
%

Our work is primarily related to studies that have been proposed to better characterize the timeliness of information in a system under different network conditions and application requirements~\cite{maatouk2020age, yin2019only, rajaraman2021not, VOI_ref}. Metrics such as Effective \ac{aoi}~\cite{yin2019only} extend the baseline definition by incorporating semantic and correctness aspects of received data, while others adapt \ac{aoi} to various queuing models, scheduling policies, and update mechanisms. Metrics like Age-Quality Information (AQI)\cite{rajaraman2021not} and Value of Information (VoI)\cite{VOI_ref} jointly optimize timeliness, data quality, and energy efficiency, often employing a utility-based function that also encompasses \ac{aoi}. Similarly, AoII metric~\cite{maatouk2020age} accounts for the correctness of updates and supports the design of optimal transmission policies under unreliable channel conditions and power constraints.
In contrast to aforementioned \ac{aoi}-based metrics, our research focuses on connecting \ac{aoi} with sustainability aspects, to present a characterization that has been overlooked in prior studies.

Our work also builds on research that explored the role of \ac{aoi} in enhancing energy efficiency~\cite{HUANG202329,xu2020info, zhang2023TWC,Hatami2021ToC, zhang2024TOC}.
Recent works have applied \ac{rl} and system-level strategies to jointly optimize \ac{aoi} and energy efficiency in diverse network settings. A deep \ac{rl} approach was proposed in~\cite{HUANG202329}, for computation offloading in \ac{iiot}, supported by queuing models. Freshness in caching-enabled \ac{iot} networks was addressed by the authors in~\cite{xu2020info},  while the proposed approach in~\cite{zhang2023TWC} leveraged \ac{mec} in a \ac{wsn} for environmental monitoring to manage energy and computational constraints. In~\cite{Hatami2021ToC}, the authors focused on energy-harvesting \ac{iot} systems with on-demand updates, while in~\cite{zhang2024TOC} the authors explored UAV-assisted data collection, revealing trade-offs between data freshness and energy cost. However, none of these works investigated the environmental impact and its connection to \ac{aoi}, as investigated in this paper.

\section{System Model}
\label{sec:minaoi_CF_b}

\begin{figure}[!t]
    \centering
    \includegraphics[width=0.5\textwidth]{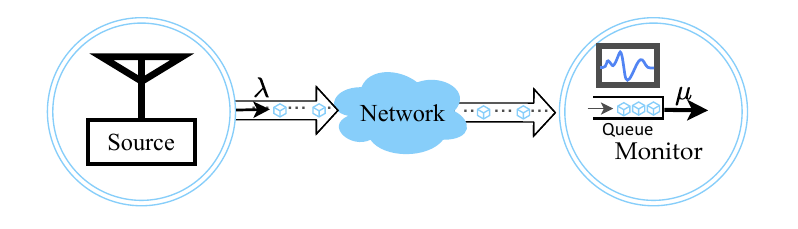}
    \caption{Illustration of the considered system.}
    \label{fig:system_model}
\end{figure}

Let us consider a basic wireless communication system consisting of an information source and a destination (sink), as depicted in Fig.~\ref{fig:system_model}. The source of information, e.g., a sensor, generates status update packets with an arrival rate of $\lambda$. These packets contain data about the measured quantity at the source, such as temperature, location, speed, as well as a timestamp indicating when the packet was generated. The status update is then transmitted over a network, which may introduce various delays due to the environment and the used access scheme. Once the status update is received at the destination, the server processes and records the most recent update with a service rate of $\mu$. The monitor uses the newly arrived update to estimate the current state of the observed phenomenon. A key performance metric in the considered system is the \ac{aoi}, denoted by $\Delta(t)$, which quantifies the freshness of the data at the monitor by measuring the time elapsed since the last received update was generated. In the $i$-th time slot, the \ac{aoi} is defined as follows:

\begin{equation}
\Delta (t) = t - u_{i}(t),
\label{eq:inst_aoi}
\end{equation}

\noindent where $u_{i}(t)$ is the time instance at which the status update was generated in the source, denoted by a timestamp.

In our work, we focus on the average \ac{aoi}, which captures the long-term behavior of information freshness in the system~\cite{kaul2012real,yates2021age}. 
It is defined as:

\begin{equation}
    \overline{\Delta}=\lim_{T \to \infty}\frac{1}{T}\int^{T}_{0}\Delta(t)dt, 
    \label{eq:avg_aoi}
\end{equation}

\noindent and serves as the main objective function of the system to minimize. Minimizing the average \ac{aoi} offers several advantages. It is less sensitive to dynamic changes, thereby it provides a more stable and representative indicator of a system’s long-term behavior. This stability makes the average \ac{aoi} especially suitable for evaluating the sustainability of communication systems, as it enables consistent assessment over longer periods, e.g., a month. Moreover, the average \ac{aoi} can be meaningfully linked to cumulative energy consumption and the resulting \ac{cf}, since frequent status updates typically entail higher power usage. 








\ac{cf} is a key metric for quantifying the environmental impact of a system~\cite{CFreport}. It is determined by two time-dependent functions: the \ac{ci} denoted by $\xi(\tau)$, and the power consumption function $P(\tau)$. \ac{ci} reflects how carbon-intensive the energy source is at a given time, as discussed in~\cite{trihinas2022towards}. We adopt the standard measure of \ac{ci} as the amount of CO$_2$-equivalent emissions produced per kilowatt-hour of energy consumed, as also discussed in Section I.  The instantaneous \ac{cf} can then be defined as the product of $\xi(\tau)$ and $P(\tau)$, thus, the cumulative \ac{cf} up to time $T$ can be expressed as follows:

\begin{equation}
    \kappa(T)=\int_{0}^{T} \xi(\tau)\,P(\tau)\,d\tau, \quad \kappa(0)=0.
\label{eq:cf_inst_def}
\end{equation}

\noindent Since both $\xi(\tau)$ and $P(\tau)$ are positive, $\kappa(\tau)$ is a monotonically increasing function. This behavior, along with standard conditions such as system stability and ergodicity, guarantees that the limit exists. Consequently, we can define the long-term average \ac{cf} ($\kappa$) as:
\begin{equation}
\kappa_{\tau} = \frac{\kappa(\tau)}{\tau}, \quad \kappa = \lim_{\tau \to \infty} \kappa_{\tau}, 
\label{eq:cf_longterm_avg}
\end{equation}

\noindent where $\kappa$ can be expressed using the average \ac{ci}, $\overline{\xi}$, and the average energy $\overline{E}$, as: 

\begin{equation}
    \kappa~[gCO_{2}eq] = \overline{\xi}~ [gCO_{2}eq/kWh]\; \overline{E}~[kWh].
    \label{eq:avg_CF_time}
\end{equation}




\section{Minimizing AoI with a CF Constraint}

We formulate an optimization problem that minimizes the average \ac{aoi}, as defined in Eq.~(\ref{eq:avg_aoi}), while constraining the cumulative average \ac{cf} (\(\overline{\kappa}\)) under a \ac{cf} budget ($K$). Our goal is to achieve a balance between system performance and limiting environmental impact in the form of \ac{cf}. We formulate the optimization problem as follows: 

\begin{equation}
\begin{aligned}
    & \min_{\lambda, \mu} \quad \overline{\Delta}(\lambda, \mu)  \\
    & \text{s.t.} \quad \overline{\kappa} \leq K, \\
    & \qquad 0 \leq \rho < 1,\\
    & \qquad 0 \leq t \leq T,
\end{aligned}
\label{eq:optimization_aoi_avg}
\end{equation}

\noindent where $\rho$ represents the system utilization and is defined as the ratio between the arrival rate and the service rate, i.e., $\rho = \frac{\lambda}{\mu}$.


We consider two queueing models: the \ac{fcfs} system, modeled as an $M/M/1$ queue, and the \ac{lcfs} system, modeled as an $M/M/1^{*}$ queue. The corresponding average \acp{aoi}, as derived in \cite{kaul2012real}, are given by:

\begin{equation}
    \overline{\Delta} = \frac{1}{\mu_{1}} \left( 1 + \frac{1}{\rho_{1}} + \frac{\rho^{2}_{1}}{1 - \rho_{1}} \right), 
    \label{eq:aaoi_MM1}
\end{equation}
and 
\begin{equation}
    \overline{\Delta}^{*} = 
    \frac{1}{\mu_{2}}+\frac{1}{\lambda_{2}},
    \label{eq:aaoi_MM1star}
\end{equation}

\noindent respectively. In addition, we consider the system-wide averages of both \ac{aoi}, \(\overline{\Delta}(\lambda, \mu)\), and cumulative \ac{cf}, \(\overline{\kappa}\), over the time interval \([0,T]\). By dividing this time period into $N$ discrete time slots, we can approximate the cumulative \ac{cf} by:

\begin{equation}
\overline{\kappa} \approx \overline{\xi}\; \overline{E}_{\mathrm{p}}\lim_{t_N \to \infty} \sum^{N}_{n=1} N_{\mathrm{p}}(t_n),
\label{eq:CF_dis}
\end{equation}

\noindent where $\overline{E}_{\mathrm{p}}$ is the average energy consumption per packet and $N_{\mathrm{p}}(t_n)$ is the number of transmitted packets at time slot $t_{n}$. Furthermore, the total number of transmitted packets can be simplified as:

\begin{equation}
\lim_{t_{N}\to \infty}\sum^{N}_{n=1} N_{\mathrm{p}}(t_n) =a \lambda t_N, 0 \leq a \leq 1,
\label{eq:N_MM1_lim}
\end{equation}

\noindent where $a$ is the probability of successful transmission. Thus, the corresponding \ac{cf} constraint is then rewritten as:

\begin{equation}
\overline{\kappa} \approx \overline{\xi}\; \overline{E}_{\mathrm{p}} a \lambda t_N \leq K.
\end{equation}

\noindent We assume $a = 1$ for both $M/M/1$ and $M/M/1^*$ models, meaning all packets are successfully transmitted. This is valid under the assumption of queuing with infinite buffers and no packet loss. However, in real-world systems with finite buffers, packet loss due to overflows can occur, leading to $a < 1$. Thus, this assumption serves as an ideal upper bound. The upper bound of arrival rate that fits into the \ac{cf} constraint is thus:


\begin{equation}
    \lambda_{\kappa} = \frac{K}{t_{N} \overline{\xi} \; \overline{E}_{\mathrm{p}}}.
    \label{eq:lambda_CF_final}
\end{equation}

The system utilization ($\rho^{\prime}$) that results in minimal average \ac{aoi} for both $M/M/1$ and $M/M/1^{*}$ models in \ac{cf} unconstrained setting are obtained by solving the corresponding equations~\cite{kaul2012real}:

\begin{equation}
\begin{aligned}
    &\rho^{4}_{1} - 2\rho^{3}_{1} + \rho^{2}_{1} - 2\rho_{1} + 1 = 0,\\
    &\rho^{2}_{2}-\rho_{2}=0.
    \label{eq:rho_poly}
\end{aligned}
\end{equation}


\noindent By solving Eq.~(\ref{eq:rho_poly}), we determine the system utilization value $\rho^{\prime}_{1} \approx 0.531$ ($\mu^{\prime}_{1} \approx 1.887\lambda^{\prime}_{1}$) that minimizes the average AoI in the $M/M/1$ model.
For the $M/M/1^{*}$ model, minimal average \ac{aoi} occurs as the arrival rate approaches the service rate, with the corresponding utilization given by:

\begin{equation}
\rho^{\prime}_{2} = 1 - \epsilon, \quad 0 < \epsilon \ll 1,
\end{equation}

\noindent where $\epsilon$ is a small positive constant. Throughout this paper, we set $\epsilon = 10^{-3}$ to approximate near-saturation conditions while avoiding system instability. Accordingly, the corresponding service rate for the $M/M/1^{*}$ model, when unconstrained by the CF budget, is given by:

\begin{equation}
\mu_2^{\prime} = \frac{\lambda^{\prime}_2}{1 - \epsilon}.
\end{equation}


On the other hand, when a \ac{cf} budget is imposed, the total number of transmitted packets is constrained, placing an upper bound on the arrival rate $\lambda$. To comply with the \ac{cf} constraint, the system must operate at a reduced transmission rate, which effectively limits the achievable average \ac{aoi}. As a result, the average \ac{aoi} for the $M/M/1$ model under the \ac{cf} constraint is given by:

\begin{equation} 
    \overline{\Delta}^{\prime}_{\kappa} = 
    \begin{cases} 
    \dfrac{1}{\mu_{1}^{\prime}} \left( 1 + \dfrac{1}{\rho_{1}^{\prime}} + \dfrac{\rho_{1}^{{\prime}^2}}{1 - \rho_{1}^{\prime}} \right), ~\text{if } \rho_{1}^{\prime} \mu_{1}^{\prime} < \lambda_{\kappa}, \\
    \dfrac{1}{\mu_{1}^{\prime}} \left( 1 + \dfrac{1}{\frac{\lambda_{\kappa}}{\mu_{1}^{\prime}}} + \frac{(\frac{\lambda_{\kappa}}{\mu_{1}^{\prime}})^2}{1 - \frac{\lambda_{\kappa}}{\mu_{1}^{\prime}}} \right), ~\text{if } \lambda_{\kappa} < \rho_{1}^{\prime} \mu_{1}^{\prime}, 
    \end{cases} 
    \label{eq:mm1_0}
\end{equation}

\noindent and for $M/M/1^{*}$ as:

\begin{equation} 
    \overline{\Delta}_{\kappa}^{*^{\prime}} \approx
    \begin{cases} 
    \frac{2}{\lambda_2}, & \quad \text{if } \rho_2^{\prime} \mu_2^{\prime} < \lambda_{\kappa}, \\
    \frac{2}{\lambda_{\kappa}}, & \quad \text{if } \lambda_{\kappa} < \rho_2^{\prime} \mu_2^{\prime}.
    \end{cases} 
    \label{eq:mm1star_0}
\end{equation}


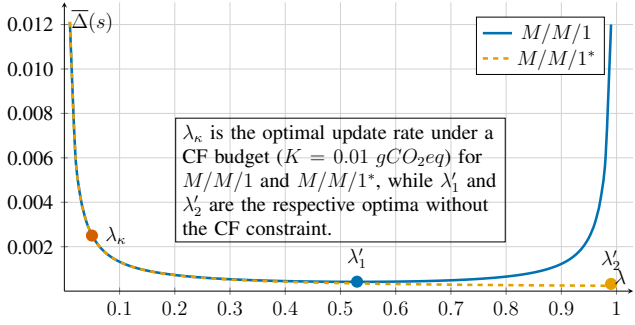
\begin{figure}
	\centering
	\large\begin{tikzpicture}
\definecolor{plotblue}{RGB}{0, 114, 178}
\definecolor{plotorange}{RGB}{230, 159, 0}

\definecolor{plotgreen}{RGB}{0, 158, 115}
\definecolor{plotpink}{RGB}{213, 94, 0}
\definecolor{plotpurple}{RGB}{204, 121, 167}
\definecolor{plotgray}{RGB}{153, 153, 153}

\begin{axis}[
	axis lines=middle,
	xlabel= $\lambda$, 
        ylabel=$\overline{\Delta} (s)$,
        scaled ticks=false,
	grid=both,
	major grid style={gray!35},
        ymin=0,ymax=0.013,
        xmin=0.0,xmax=1.03,
        ytick ={0, 0.002,0.004,0.006,0.008,0.010, 0.012 },
        yticklabels = {0, 0.002,0.004,0.006,0.008,0.010, 0.012 },
	xtick ={0.1, 0.2, 0.3, 0.4, 0.5, 0.6, 0.7, 0.8, 0.9, 1.0},
	width=1.5\linewidth,
	height=0.85\linewidth,
	legend style={at={(0.95,0.95)}},
]

\addplot[line width=1.5pt, color=plotblue,  smooth, mark options={solid}]table[x=rho,y=AoI_MM1,col sep=comma]{csv_data/AoI_vs_Rho.csv};

\addplot[line width=1.5pt, color=plotorange,  dashed, smooth, mark options={solid}]table[x=rho,y=AoI_MM1_star,col sep=comma]{csv_data/AoI_vs_Rho.csv};

\node[circle, fill=plotblue, inner sep=2.5pt, label=above:$\lambda^{\prime}_{1}$] at (53, 4.25) {};
\node[circle, fill=plotorange, inner sep=2.5pt, label=above:$\lambda^{\prime}_{2}$] at (99, 3.25) {};
\node[circle, fill=plotpink, inner sep=2.5pt, label=right:$\lambda_{\kappa}$] at (5,25) {};
 
\node[draw, text width=6.5cm] at (50,50) {$\lambda_{\kappa}$ is the optimal update rate under a CF budget ($K= 0.01~gCO_{2}eq$) for $M/M/1$ and $M/M/1^{*}$, while $\lambda^{\prime}_{1}$ and $\lambda^{\prime}_{2}$ are the respective optima without the CF constraint.};







\addlegendentry{$M/M/1$};
\addlegendentry{$M/M/1^{*}$};
\addlegendentry{$\lambda^{\prime}_{2},~\text{without}~\overline{\kappa}\leq K$};
\addlegendentry{$\lambda^{\prime}_{2},~\text{with}~\overline{\kappa}\leq K$};

\end{axis}

\end{tikzpicture}
         \caption{\ac{aoi} as a function of arrival rate ($\lambda$) for $M/M/1$ and $M/M/1^*$ models.}
	\label{fig:aoivsrho}
	\vspace{-10pt}
\end{figure}


Fig.~\ref{fig:aoivsrho} illustrates the relationship between the arrival rate $\lambda$ and the average \ac{aoi} for the $M/M/1$ and $M/M/1^*$ models, evaluated under both unconstrained and \ac{cf}-constrained conditions. The system parameters used in this analysis are provided in Table~\ref{tab:system_parameters}.

\begin{table}[t]
    \centering
    \caption{Summary of System Parameters.}
    \begin{tabular}{llc}
        \toprule
        \textbf{Parameter} & \textbf{Description} & \textbf{Value} \\
        \midrule
        \( \overline{\xi} \) & Average CI in Slovenia (2024) & $198$ ~$\frac{gCO_{2}eq}{kWh}$ \\
        \( P_{\mathrm{T}} \) & Transmitting Power & $1$~$W$ \\
        \( R\) & Transmitting Rate & \( 10^8 \) $\frac{bits}{s}$ \\
        \( \text{MTU} \) & Maximum Transmission Unit & 12,000 bits \\
        \( T_{\text{p}} \) & Transmission Time per Packet & \( \frac{\text{MTU}}{R} \) $s$ \\
        \( t_N \) & Time Period & $3600$~$s$ (1 hour) \\
        \( \overline{E}_{\mathrm{p}} \) & Average Energy per Packet & \( P_{\mathrm{T}} \times T_{\text{p}} \)~$J$ \\
        \( K \) & Carbon Footprint Budget & \( 5 \times 10^{-2} \) $gCO_{2}eq$ \\
        \( B \) & Bandwidth & \( 1 \times 10^{6} \)~$Hz$ \\
        \( |h|^{2} \) & Channel Gain & \( 1 \)   \\
        \( \sigma^{2} \) & Noise Power & \( 10^{-4} \)~$W$  \\
        \bottomrule
    \end{tabular}
    \vspace{-15pt}
    \label{tab:system_parameters}
\end{table}

For the $M/M/1$ model (represented by the blue solid curve), the average \ac{aoi} ($\overline{\Delta}$) exhibits a characteristic U-shaped curve with high \ac{aoi} at low and high $\lambda$ due to infrequent updates and queuing delays, respectively. 
In contrast, the \ac{aoi} in the $M/M/1^*$ model (shown as orange dashed line), decreases monotonically with increasing $\lambda_{2}$, achieving its minimum at near-saturation ($\rho^{\prime}_2 = 0.99$), due to its priority on processing the most recent updates.
However, when, for example, a strict \ac{cf} budget of $0.01$~$gCO_{2}eq$, both models must reduce their transmission rates to comply with the environmental constraint. Under this \ac{cf} budget, the maximum allowable arrival rate is limited to  $\lambda_{\kappa} \approx 0.05$. Moreover, this convergence indicates that, under a \ac{cf} constraint, the inherent differences between the \ac{fcfs} ($M/M/1$) and \ac{lcfs} ($M/M/1^*$) strategies become less significant, since both are primarily constrained by the arrival rate. This observation highlights the trade-off between maintaining data freshness (minimizing the average \ac{aoi}) and achieving sustainability objectives.

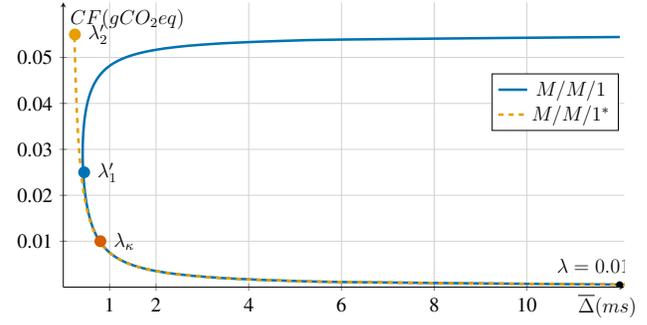
\begin{figure}[t!]
	\centering
	\large \begin{tikzpicture}
\definecolor{plotblue}{RGB}{0, 114, 178}
\definecolor{plotorange}{RGB}{230, 159, 0}

\definecolor{plotgreen}{RGB}{0, 158, 115}
\definecolor{plotpink}{RGB}{213, 94, 0}
\definecolor{plotpurple}{RGB}{204, 121, 167}
\definecolor{plotgray}{RGB}{153, 153, 153}

\begin{axis}[
	axis lines=middle,
	xlabel= $\overline{\Delta}(ms)$,
        ylabel=$CF(gCO_2eq)$,
        scaled ticks=false,
	grid=both,
	major grid style={gray!35},
        ymin=0,
        xmin=0,
        xmax=.0121,
        ymax=0.062,
	xtick ={0.001, 0.002,0.004,0.006,0.008,0.010},
        xticklabels = {1,2,4,6,8,10 },
        ytick={0.01,0.02,0.03,0.04,0.05}, 
        yticklabels={0.01,0.02,0.03,0.04,0.05},
         xlabel style={at={(ticklabel* cs:1)}, xshift=-10.0pt, anchor=north},
	width=1.5\linewidth,
	height=0.85\linewidth,
	legend style={at={(0.99,0.75)}},
]

 
\addplot[line width=1.5pt, color=plotblue,  smooth, mark options={solid}]table[x=AoI_MM1,y=Carbon_Footprint,col sep=comma]{csv_data/Combined_AoI_CF_Data.csv};

\addplot[line width=1.5pt, color=plotorange,  dashed, smooth, mark options={solid}] table[x=AoI_MM1star,y=Carbon_Footprint,col sep=comma]{csv_data/Combined_AoI_CF_Data.csv};

\node[circle, fill=plotblue, inner sep=2.5pt, label=right:$\lambda^{\prime}_{1}$] at (4.5, 250) {};
\node[circle, fill=plotorange, inner sep=2.5pt, label=right:$\lambda^{\prime}_{2}$] at (2.5, 550) {};

\node[circle, fill=plotpink, inner sep=2.5pt, label=right:$\lambda_{\kappa}$] at (8,100) {};

\node[circle, fill=black, inner sep=1.5pt, ] at (120, 5.0) {};
\node[label={above:$\lambda=0.01$}, xshift=-21.0pt, anchor=north] at (122, 15.5) {};

\addlegendentry{$M/M/1$};
\addlegendentry{$M/M/1^{*}$};


\end{axis}

\end{tikzpicture}
         \caption{\ac{aoi} versus \ac{cf} for the \(M/M/1\) and \(M/M/1^*\) models.}
	\label{fig:aoivsCF_0}
    \vspace{-10pt}
\end{figure}

In Fig.~\ref{fig:aoivsCF_0}, we present the average \ac{aoi} as a function of the \ac{cf} for the two queuing models. The figure illustrates the relationship between \ac{aoi} and \ac{cf}, highlighting the inherent performance trade-offs. Interestingly, when the update rate is low (e.g., $\lambda = 0.01$), the system exhibits a high \ac{aoi} but a very low \ac{cf}. As the update rate increases, the \ac{cf} also increases, while, as expected, the \ac{aoi} decreases for both queuing models.
The \ac{aoi} continues to decrease until reaching the optimal update rate for the $M/M/1$ model, denoted as $\lambda_1^{'}$. Beyond this point, for the $M/M/1^{*}$ model, further increases in the update rate lead to a significantly higher \ac{cf} with only marginal gains in reducing the \ac{aoi}. Conversely, for the $M/M/1$ model, the average \ac{aoi} begins to increase while the \ac{cf} continues to rise. Consequently, from a sustainability perspective, when the $M/M/1$ model is employed, the most beneficial update rate is $\lambda = 0.53$.
Furthermore, we observe that for the $M/M/1^{*}$ model, a twofold increase in \ac{cf} does not result in a proportional reduction in the average \ac{aoi}. For example, when the \ac{cf} increases from $0.02$~$gCO_{2}eq$ to $0.04$~$gCO_{2}eq$, the corresponding decrease in the average \ac{aoi} is minimal for $M/M/1^{*}$ model. Hence, we can conclude that the relationship between \ac{aoi} and \ac{cf} is nonlinear. Moreover, the \ac{ci} also influences this relationship, as analyzed in more detail in the following section.

\section{
Analysing Tradeoff under Dynamic CIs}
\label{sec:minaoi_sus}

Building on trade-off between minimizing \ac{aoi} and managing environmental impact, we now examine how the sustainability constraint interacts with varying \ac{ci}. While the \ac{cf} budget limits energy consumption, the \ac{ci} further shapes the feasible operating region and the achievable average \ac{aoi}, as formalized in Eq.~(\ref{eq:avg_CF_time}). To capture this effect, we solve the \ac{aoi}-minimization problem for each monthly \ac{ci} realization (Fig.~\ref{fig:intro}) under two edge cases: (i) sustainability-driven operation, where transmission power ($P_{\mathrm{T}}$) is constrained, and (ii) \ac{qos}-driven operation, where a minimum \ac{snr} is enforced at the cost of higher energy consumption.




\subsection{AoI Minimization under CF and power constraints}


We extend the discrete-time AoI minimization problem in Eq.~(\ref{eq:optimization_aoi_avg}) by incorporating both a transmitting power constraint and a monthly \ac{ci} profile indexed by $i$, shown in Fig.~\ref{fig:intro}. The extended problem is formulated as:

\begin{equation}
\begin{aligned}
    & \min_{\lambda_i,\mu_i} \quad \overline{\Delta}_i\left( \lambda_i,\mu_i\right) \\
    & \text{s.t.} \quad \overline{\xi_i}\; \overline{E}_{\mathrm{p}} \lambda t_N \leq K, \forall i =1,\ldots, 12,\\
    & \qquad P_{\mathrm{T}}\leq P_{\max},\\
    & \qquad 0 \leq \rho_i < 1,\\
    & \qquad 0 \leq t \leq t_N,
\end{aligned}
\label{eq:optimization_aoi_avg_PTX}
\end{equation}

\noindent where $P_{\mathrm{T}}$ is the transmitting power and $P_{\max}$ is its maximum allowable value. 
The \ac{cf} constraint is derived from the combination of transmitting power and the monthly \ac{ci}, leading to a bound on the arrival rate:

\begin{equation}
    \lambda_{\mathrm{P,max},i}=\frac{K}{\overline{\xi_i}P_{\mathrm{T,max}}T_{\mathrm{p}}t_{N}},
    \label{eq:lambda_p_max}
\end{equation}

\noindent where $T_{\mathrm{p}}$ is the transmission time per packet, calculated as the packet size (including overhead) divided by the data rate. Using this constraint, the average \ac{aoi} for both the $M/M/1$ and $M/M/1^{*}$ models can be computed by applying Eqs.~(\ref{eq:mm1_0}) and~(\ref{eq:mm1star_0}) and substituting $\lambda_{\mathrm{P,max},i}$ for $\lambda_{\kappa}$. 

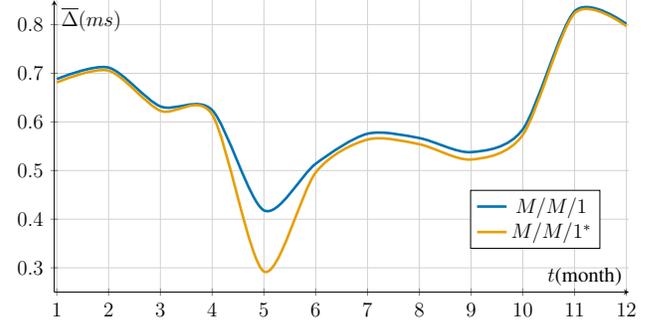
\begin{figure}[t!]
    \centering
    \large \begin{tikzpicture}
\definecolor{plotblue}{RGB}{0, 114, 178}
\definecolor{plotorange}{RGB}{230, 159, 0}

\definecolor{plotgreen}{RGB}{0, 158, 115}
\definecolor{plotpink}{RGB}{213, 94, 0}
\definecolor{plotpurple}{RGB}{204, 121, 167}
\definecolor{plotgray}{RGB}{153, 153, 153}

\begin{axis}[
	axis lines=middle,
	xlabel= $t$(month), 
        ylabel=$\overline{\Delta} (ms)$ ,
        scaled ticks=false,
	grid=both,
	major grid style={gray!35},
        ymin=0.00025,
        ymax=0.00085,
        ytick ={0.0003,0.0004,0.0005,0.0006,0.0007,0.0008},
        yticklabels = {$0.3$,$0.4$,$0.5$,$0.6$,$0.7$, $0.8$},
        xmin=0.95,
        xmax=12.05,
	xtick={1,2,3,4,5,6,7,8,9,10,11,12},
	width=1.5\linewidth,
	height=0.85\linewidth,
	legend style={at={(0.95,0.35)}},
]

 
\addplot[line width=1.5pt, color=plotblue,  smooth, mark options={solid}]table[x=month,y=MM1,col sep=comma]{csv_data/Clean_AOI_12_months_table.csv};

\addplot[line width=1.5pt, color=plotorange, smooth, smooth, mark options={solid}]table[x=month,y=MM1Star,col sep=comma]{csv_data/Clean_AOI_12_months_table.csv};



\addlegendentry{$M/M/1$ };
\addlegendentry{$M/M/1^{*}$ };

\end{axis}

\end{tikzpicture}
    \caption{Average \ac{aoi} over monthly CI for $M/M/1$ (blue line) and $M/M/1^*$ (orange line) models.}
    \label{fig:aoi_pt_12}
    
\end{figure}

Fig.~\ref{fig:aoi_pt_12} shows how month-to-month variations in \ac{ci} affect the achievable \ac{aoi} under this fixed \ac{cf} budget. Note that a \ac{cf} budget of $0.05$~$gCO_{2}eq$ is assumed in the  analysis.
A lower \ac{ci} allows for higher system utilization ($\rho$), thereby reducing average \ac{aoi}. This effect is more pronounced in the $M/M/1^{*}$ model, which continues to benefit as $\rho$ approaches 1, while the $M/M/1$ model reaches its minimal AoI near $\rho \approx 0.531$. In contrast, high \ac{ci} values constrain both models to lower $\rho$, resulting in elevated AoI levels. For example, in May, the minimal AoI for $M/M/1$ is $0.42$~ms, approximately 1.45 times higher than that of $M/M/1^{*}$ at $0.29$~ms. In contrast, in November, both models perform similarly, with AoI values of $0.827$~ms and $0.823$~ms, respectively.

\begin{figure}[t]
    \centering
        \begin{subfigure}[h]{0.49\columnwidth}
        \includegraphics[width=\linewidth]{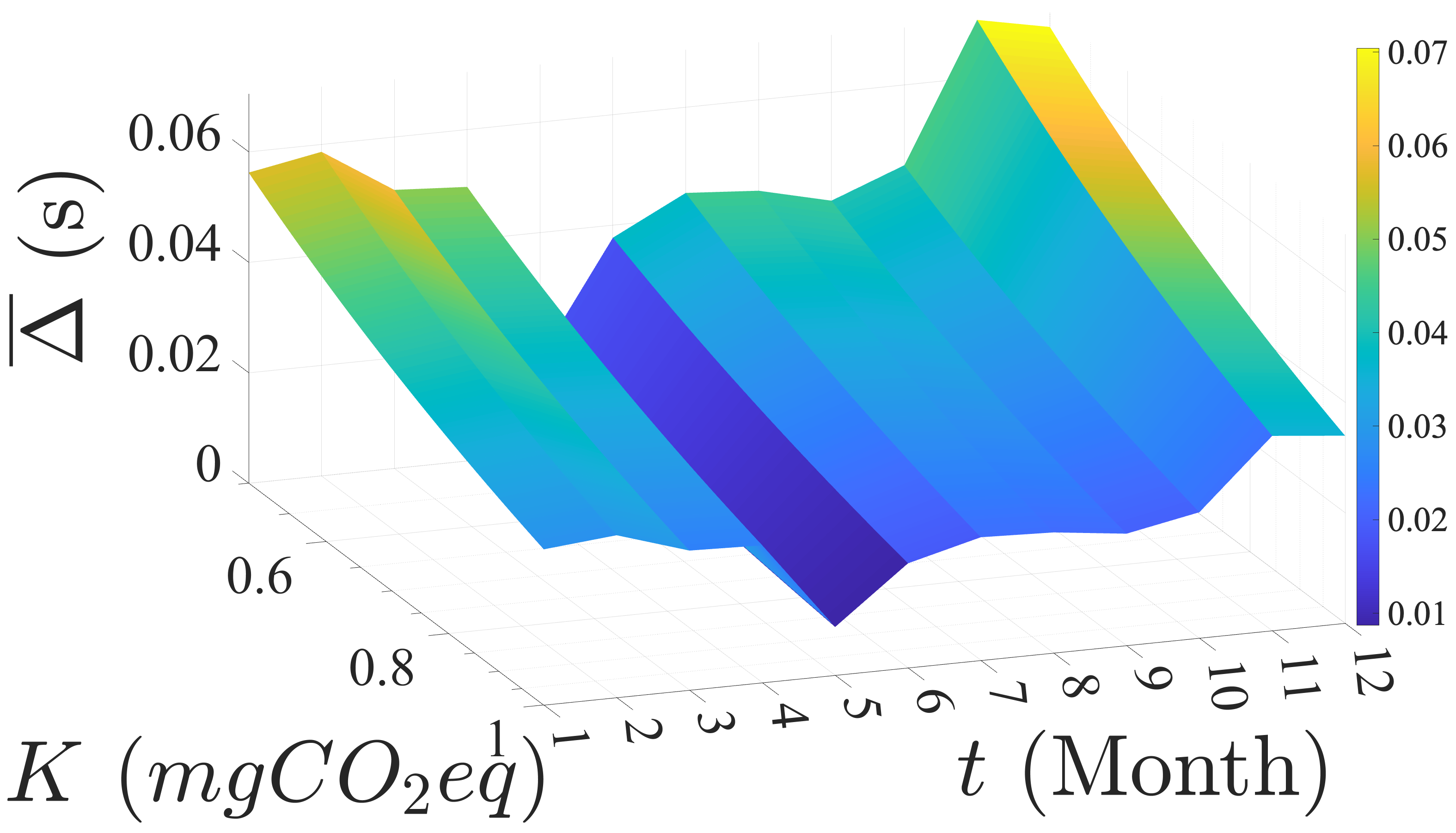}

        \caption{AoI for $M/M/1$.}
        \label{fig:AoI_pt_3D_mm1}
    \end{subfigure}
    \begin{subfigure}[h]{0.49\columnwidth}
        \includegraphics[width=\linewidth]{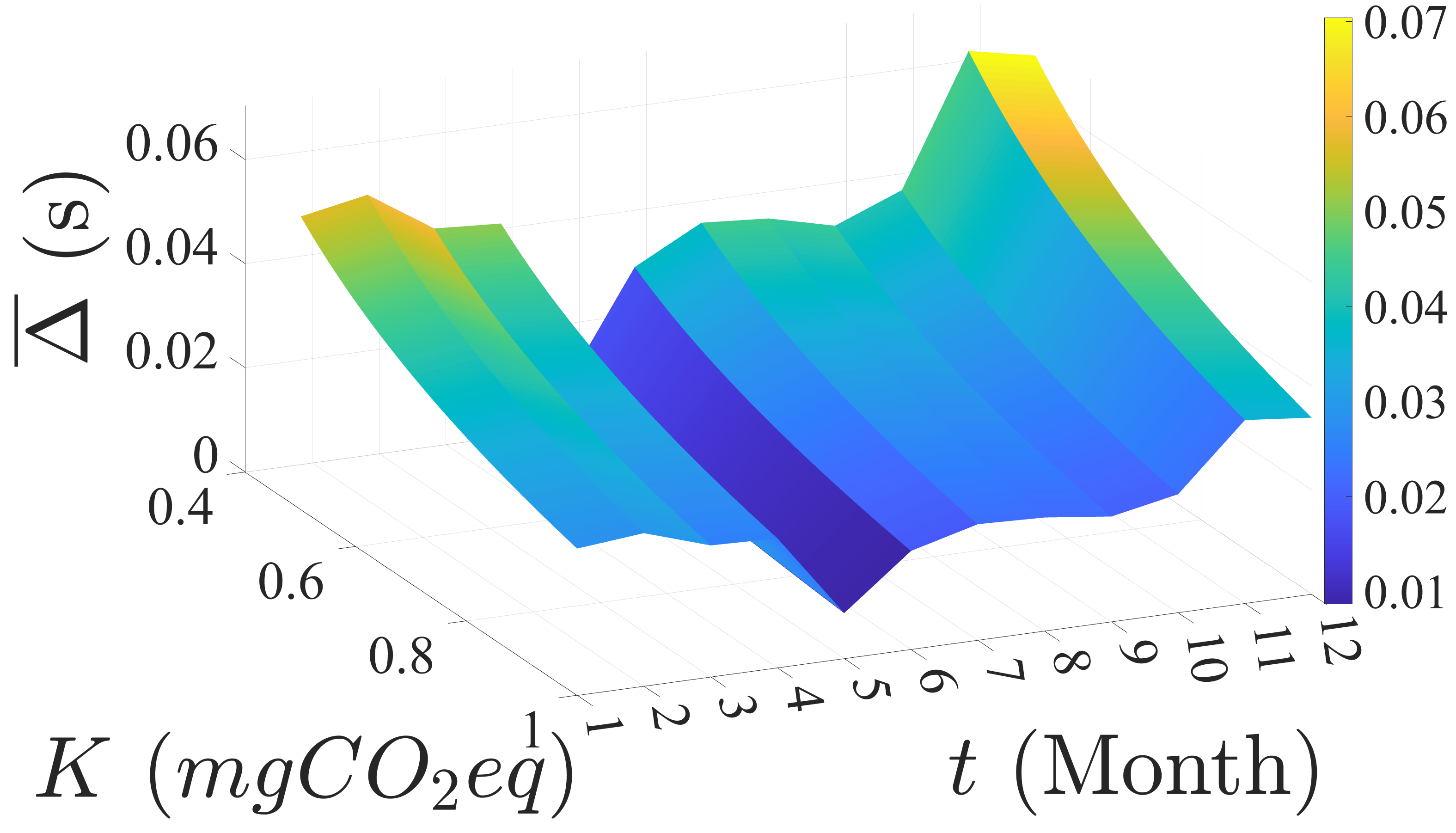}

        \caption{AoI for $M/M/1^*$.}
        \label{fig:AoI_pt_3D_mm1_star}
    \end{subfigure}

    \caption{Average AoI over time ($x$-axis) versus \ac{cf} budget ($K$) from $0.5$ to $1$ $mgCO_{2}eq$ under identical transmit power ($P{\max}$) constraints.} 
    \label{fig:AoI_CF_PT_3D}
    \vspace{-10pt}
\end{figure}

In addition, we vary the \ac{cf} budget from $0.5$~$mgCO_{2}eq$ to $1$~$mgCO_{2}eq$. As shown in Fig.~\ref{fig:AoI_CF_PT_3D}, the month-to-month behavior of \ac{aoi} confirms the results in Fig.~\ref{fig:aoi_pt_12}. The optimal \ac{aoi} ($z$-axis) decreases monotonically with K for each month. That is because a larger \ac{cf} ($K$) relaxes the upper bound of the arrival rate ($\lambda_{\mathrm{P},\max,i}$), whereas a smaller $K$ (tighter bound) causes the surfaces to flatten.





\subsection{AoI Minimization under CF budget and QoS constraint}
Next, we aim to minimize the average \ac{aoi} in $M/M/1$ and $M/M/1^{*}$ models while satisfying \ac{cf} and \ac{qos} constraints. In particular, we impose a lower bound on the \ac{snr} to ensure a reliable link between the source and the sink. For the monthly analysis, we assume the wireless channel is quasi-static, and has steady \ac{snr} over the period. The optimization problem is then expressed as follows:

\begin{equation}
\begin{aligned}
    & \min_{\lambda_{i},\mu_{i}} \quad \overline{\Delta}_{i}\left( \lambda_{i}, \mu_{i} \right) \\
    & \text{s.t.} \quad \overline{\xi}_{i}\; \overline{E}_{\mathrm{p}} \lambda t_N \leq K, \\
    & \qquad 
    SNR\geq SNR_{\min},\\
    & \qquad  0 \leq \rho < 1,\\
    & \qquad 0 \leq t \leq T,
\end{aligned}
\label{eq:optimization_aoi_avg_SNR}
\end{equation}

\noindent where $SNR_{\min}$ denotes the minimum required \ac{snr} for a successful transmission at the minimum achievable data rate $R$. This relationship is expressed as:

\begin{equation}
\begin{aligned}
    R \geq R_{\min}&= B\log_{2}\left(1+SNR_{\min}\right) \\
             &= B\log_{2}\left(1+\frac{P_{\mathrm{T,\min}}|h|^2}{\sigma^{2}}\right),
\end{aligned}
\label{eq:rate_ineq}
\end{equation}

\noindent where $B$ is the bandwidth, $|h|^2$ is the channel gain, and $\sigma^2$ is the noise power. This \ac{snr} constraint implies a minimum transmit power ($P_{\mathrm{T},\min}$). By substituting $P_{\mathrm{T},\min}$ into the \ac{cf} constraint from Eq.~(\ref{eq:optimization_aoi_avg_SNR}), we obtain an upper bound for the arrival rate ($\lambda_{\mathrm{QoS,max}}$), which can be expressed as: 

\begin{equation}
    \lambda_{\mathrm{QoS,max},i}=\frac{K|h|^{2}}{\overline{\xi}_{i}SNR_{\min}\sigma^{2}T_{\mathrm{p}}t_{N}}.
    \label{eq:lambda_SNR_max}
\end{equation}

\noindent This ensures that the system maintains a guaranteed level of \ac{qos} by enforcing a lower bound on transmission reliability and throughput with respect to the targeted \ac{cf} budget ($K$). As a result, the optimal solutions for the average \ac{aoi} under \ac{cf} and \ac{qos} constraints in the $M/M/1$ and $M/M/1^{*}$ models can be obtained by replacing the upper bound $\lambda_{\kappa}$ in Eqs.~(\ref{eq:mm1_0}) and~(\ref{eq:mm1star_0}) with $\lambda_{\mathrm{QoS,max}}$ (see Eq.~(\ref{eq:lambda_SNR_max})).

\begin{figure}[t!]
    \centering
        \begin{subfigure}[h]{0.49\columnwidth}
        \includegraphics[width=\linewidth]{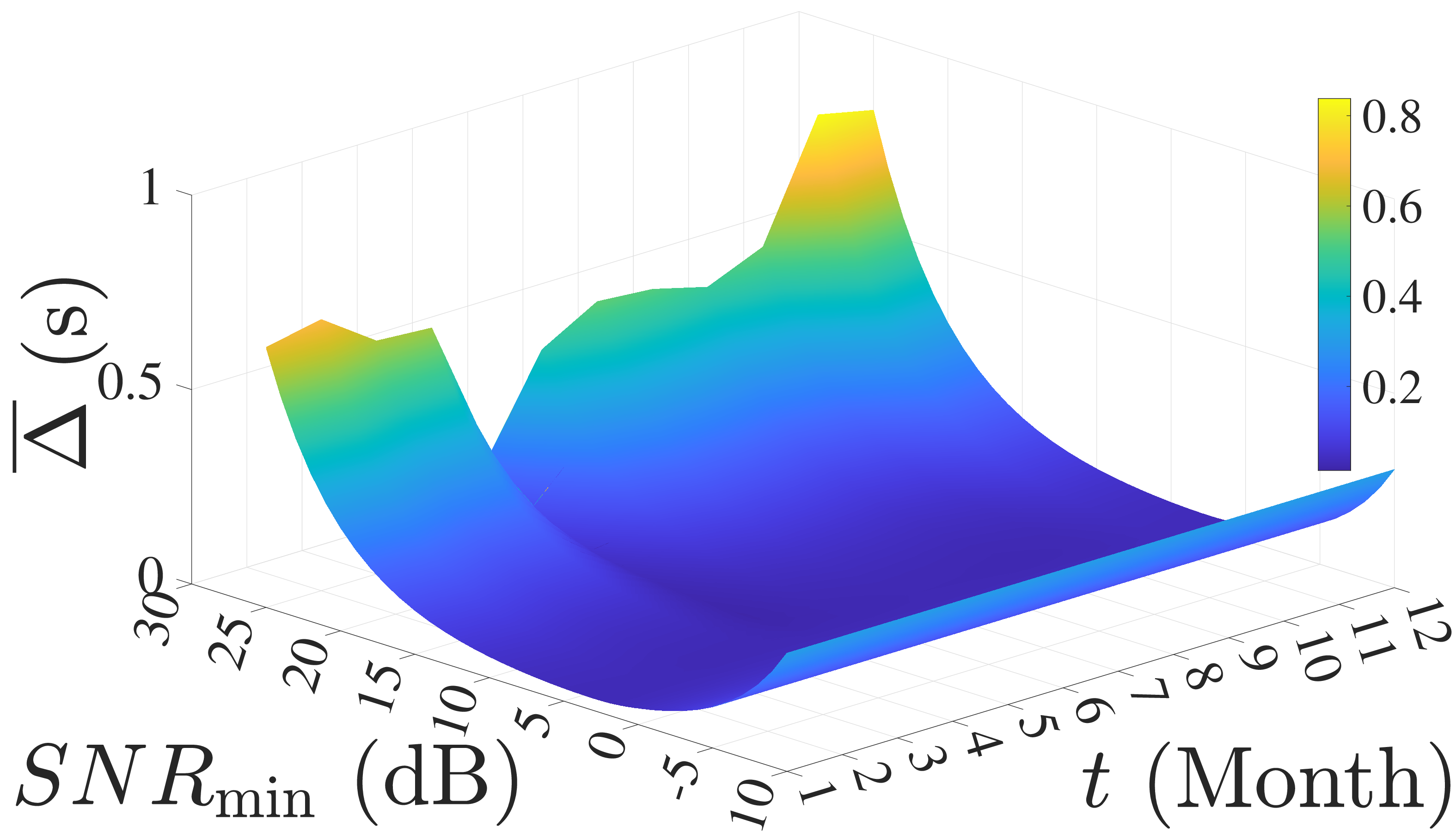}

        \caption{AoI for $M/M/1$.}
        \label{fig:aoi_QOS_MM1}
    \end{subfigure}
    \begin{subfigure}[h]{0.49\columnwidth}
        \includegraphics[width=\linewidth]{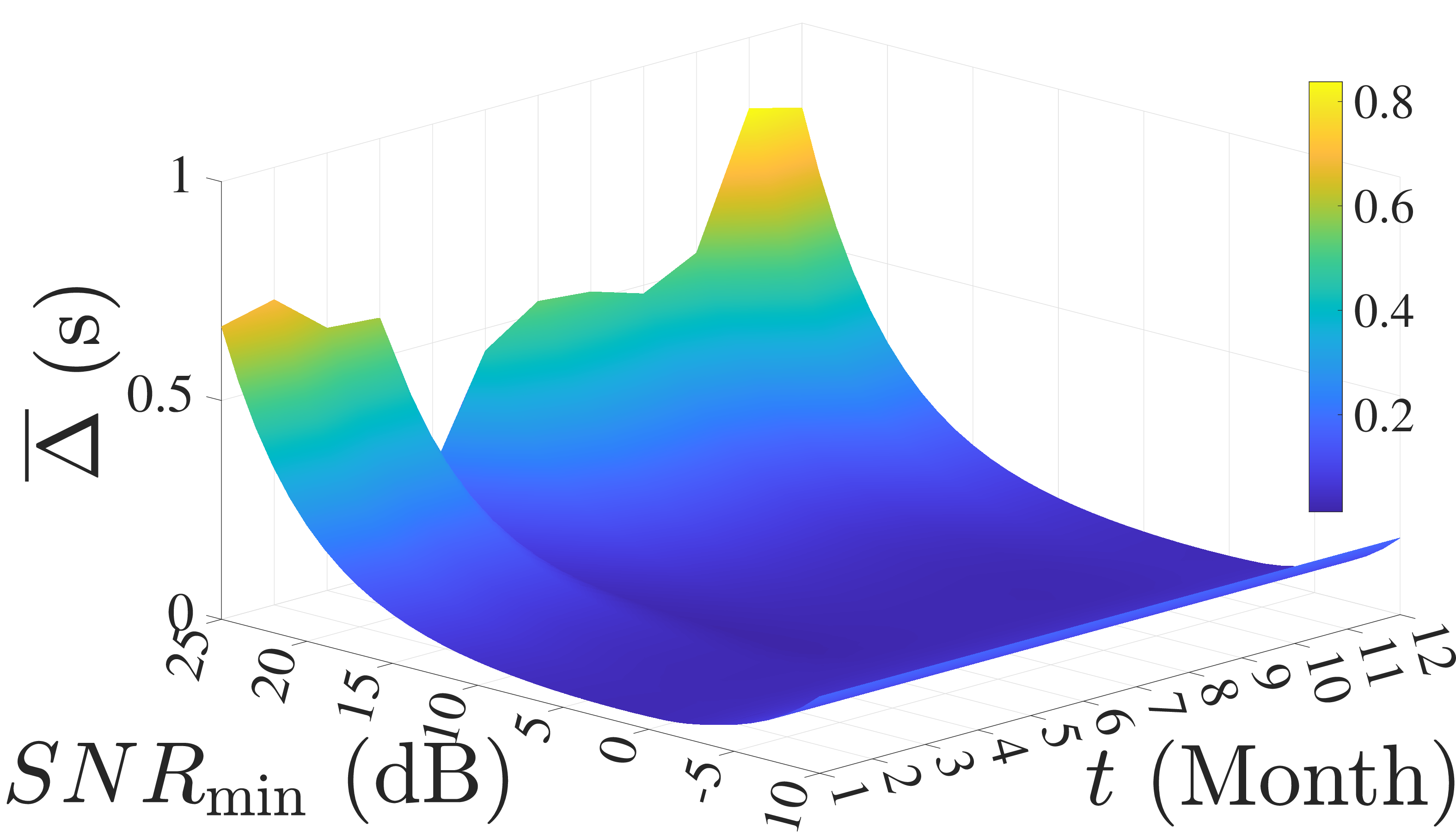}

        \caption{AoI for $M/M/1^*$.}
        \label{fig:aoi_QOS_MM1_star}
    \end{subfigure}

    \caption{Average AoI across months ($x$-axis) versus \ac{snr} budget ($SNR_{\min}$) constraint from $-10$ to $30$~$dB$ under same \ac{cf} ($K=0.016$~$\mu g$) constraint.} 
    \label{fig:aoi_qos_all}
    \vspace{-15pt}
\end{figure}

We now analyze how the average \ac{aoi} is influenced by dynamic \ac{ci} and \ac{qos} requirements under a \ac{cf} budget. The system parameters are set as follows: system utilization $\rho = 0.5$, noise power $\sigma^{2} = 10^{-4}$~$W$, minimum required signal-to-noise ratio $SNR_{\min} = 5$~$dB$, and \ac{cf} budget $K = 0.016~gCO_{2}eq$. The results for the $M/M/1$ and $M/M/1^{*}$  models are shown in Fig.~\ref{fig:aoi_QOS_MM1} and Fig.~\ref{fig:aoi_QOS_MM1_star}, respectively. These results show that CI variability affects the achievable average \ac{aoi} across \ac{snr} levels; under a fixed \ac{cf} budget, \ac{aoi} is U-shaped in \ac{snr}\textsubscript{min}, therefore, decreasing at moderate \ac{snr} and rising again as power constraints tighten. Furthermore, under a fixed \ac{cf} budget, increasing $SNR_{\min}$ improves the achievable arrival rate, thereby reducing the average \ac{aoi}. For example, as $SNR_{\min}$ increases from $-10$ to $5$~$dB$, the average \ac{aoi} drops from 0.3 to 0.1~ms in the $M/M/1$ model, and from 0.2 to 0.1~ms in the $M/M/1^{*}$ model.

As $SNR_{\min}$ continues to increase, the required transmit power $P_{\mathrm{T},\min}$ also grows, eventually reaching the set \ac{cf} budget. Consequentially, this leads to a tighter bound on the allowable arrival rate (see Eq.~\ref{eq:lambda_SNR_max}), which in turn causes the average \ac{aoi} to rise again in both models and resulting in a U-shaped \ac{aoi} curve, similarly as we have shown in Fig.~\ref{fig:aoivsCF_0}. Additionally, monthly variations in \ac{ci} significantly affect the average \ac{aoi}. For example, at $SNR_{\min} = 30$~dB, the month with the highest \ac{ci} yields an \ac{aoi} nearly four times higher than that of the month with the lowest CI, for both queuing models. These results highlight the complex trade-offs between data freshness, signal quality, and environmental sustainability, and highlight the importance of system-level optimization in meeting stringent QoS targets without exceeding environmental constraints.

\section{Conclusion and future work}
\label{sec:conclusion}

This paper examined the trade-off between information freshness and environmental impact by incorporating the \ac{cf} of transmitted updates into \ac{aoi} framework. We derived closed-form expressions for average \ac{aoi} under \ac{cf} constraints in $M/M/1$ and $M/M/1^{\ast}$ models, showing that sustainability requirements significantly alter optimal system utilization. Our results show that as update rates increase, improvements in \ac{aoi} diminish while \ac{cf} grows sharply, confirming a nonlinear trade-off between timeliness and sustainability. Extending the analysis to time-varying \ac{ci}, we showed that monthly fluctuations in \ac{ci} significantly influence achievable \ac{aoi} and that sustainable operation depends on balancing transmission power, \ac{cf} budgets, and \ac{qos} requirements. High \ac{ci} conditions restrict achievable \ac{aoi}, while lower \ac{ci} periods enable greener and more frequent updates. 
Additionally, the two analyzed use cases further demonstrated how transmission power, \ac{qos} constraints, and \ac{ci} affect the \ac{aoi}–\ac{cf} trade-off. 

Our future work will build on the presented trade-offs by focusing on the design of real-time, adaptive data collection mechanisms that jointly account for \ac{aoi}, \ac{cf} budgets, and \ac{qos} requirements. Moreover, we will exploit the seasonality of \ac{ci} to schedule tasks during greener periods and apply predictive models to anticipate data staleness, thereby enabling proactive updates when timeliness is critical or update costs are high. The ultimate goal is to develop scalable, \ac{cf}-aware strategies for 6G and large-scale \ac{iot} systems that balance sustainability, information freshness, and \ac{qos}.

\section*{Acknowledgements}
This work was supported by the 
Slovenian Research Agency under grants P2-0016, MN-0009-106, and J2-50071.

\balance

\bibliographystyle{IEEEtran}
\bibliography{bibliography}

\end{document}